Biological Sciences: Biophysics

# Simulation, Experiment, and Evolution: Understanding Nucleation in Protein S6 Folding

Isaac A. Hubner[†], Mikael Oliveberg[‡], and Eugene I. Shakhnovich[†*]

[†]Department of Chemistry and Chemical Biology
Harvard University
12 Oxford Street
Cambridge, MA 02138

[‡]Department of Biochemistry
Umeå University, S-901 87
Umeå, Sweden

[*]corresponding author
tel: 617-495-4130
fax: 617-384-9228
email: eugene@belok.harvard.edu

**Abstract**

In this study, we explore nucleation and the transition state ensemble of the ribosomal protein S6 using a Monte Carlo Go model in conjunction with restraints from experiment. The results are analyzed in the context of extensive experimental and evolutionary data. The roles of individual residues in the folding nucleus are identified and the order of events in the S6 folding mechanism is explored in detail. Interpretation of our results agrees with, and extends the utility of, experiments that shift $\phi$-values by modulating denaturant concentration and presents strong evidence for the realism of the mechanistic details in our Monte Carlo Go model and the structural interpretation of experimental $\phi$-values. We also observe plasticity in the contacts of the hydrophobic core that support the specific nucleus. For S6, which binds to RNA and protein after folding, this plasticity may result from the conformational flexibility required to achieve biological function. These results present a theoretical and conceptual picture that is relevant in understanding the mechanism of nucleation in protein folding.

Understanding the transition state (TS) is among the major technical and intellectual milestones towards understanding the protein folding reaction (1). Several recent studies (2-6) have attempted to construct transition-state ensemble (TSE) structures by utilizing $\phi$-values as structural restraints in unfolding simulations. Through extensive studies of the experimentally and computationally benchmarked protein G (7), we have shown that experimental $\phi$-values ($\phi^{exp}$) may be employed in simulation to construct a putative TSE, but that measurement of a conformation's transmission coefficient ("probability to fold", $p_{fold}$) is the only means by which a structure may be classified as a member of the TSE. However, one must also be cautious in choosing which $\phi^{exp}$ to use since the point mutations on which they are based may alter protein stability or structural features of the TSE, making normalization to the wild-type data ambiguous (8). Given that our method for studying the structure of the TSE has been validated in the complicated case of protein G folding (7), we are now able to carry out such analysis for other proteins on a comparative basis to aid in the distinction between experimental inconsistencies, noise, or artifacts and to determine the common denominators of the critical nucleus in protein folding.

The split β-α-β ribosomal protein S6 from *Thermus thermophilus* consists of 97 residues in a four-stranded β-sheet packed against two α-helices with a hydrophobic core (9). Functionally, S6 binds to both RNA and its protein partner S18 in a cooperative manner during the intermediate stage of 30S ribosomal subunit formation (10). S6 is an ideal candidate for computational study, due to the large body of high-quality experimental data available, including extensive kinetic and $\phi$-value data at varying denaturant concentrations (11), circular-permutant studies that reflect rearrangements in the TSE (12), and studies of salt-induced off-pathway intermediates (13). Detailed structural information also exists for the function of S6 (10).

We use simulation to structurally interpret the combined set of S6 $\phi$-values. We begin by generating an ensemble of structures consistent with structural restraints based on $\phi^{exp}$. After characterizing each ensemble conformation by measuring its $p_{fold}$, we construct a detailed model of the TSE and the events occurring before and after nucleation. This formalized treatment of $\phi^{exp}$ allows microscopic analysis and reconstruction of the folding nucleation process. Our results support the idea that the experimentally observed denaturant-induced shifts in $\phi$-values shift the TSE along the free-energy profile and hence probe events earlier or later along the folding pathway (as interpreted through the Hammond postulate). We also analyze residue conservation patterns of S6 to determine the evolutionary history of nucleus residues in the split β-α-β family. Through a combination of experimental data, evolutionary information, and all-atom simulation we are able to extend interpretation of experimental data and create unified and ordered, atomic level description of nucleation, the transition state and folding in the S6 protein.

**Theory and Methods**
**Model system.** Our protocol was previously implemented for reconstructing the TSE of CI2 (6) and protein G (7) and has been used to simulate the complete folding of protein G (14) and crambin (15) from random coil to < 1Å distance RMS (dRMS) from the native state. The model includes a hard sphere representation of all non-hydrogen atoms in the backbone and side chains, a full representation of backbone and side chain rotational degrees of freedom, a square well Go potential (16) with native contacts having a -1 attraction and non-native contacts having a +1 repulsion, and a MC move set (with localized backbone and side chain moves) that maintains chain connectivity, planar peptide bonds, and excluded volume at each step. As a measure of structural similarity to the native state, dRMS is computed as $\sqrt{\langle (D-D_0)^2 \rangle}$ where $D$ and $D_0$ are pairwise $C_\alpha$ distances in a selected and native conformation.

This model has the advantage of allowing for a statistically significant number of trajectories to be collected while including atomic-resolution details, such as side chain packing. Go potentials adequately represent the thermodynamics and kinetics of proteins with minimal energetic frustration and allow the complete folding process to be studied (17). Their use is also motivated by the theoretical and experimental finding that transition state is robust with respect to selection of specific sequences that fold to a given structure and potentials sets used to design and fold sequences (18-20). Go models have been used to propose folding mechanisms (21), predict folding rates, and interpret $\phi$-values (22-24). They have also successfully predicted $\phi$-values for several proteins (25-27). Presently, there is no general potential capable of folding $\alpha/\beta$ proteins so Go potentials present the best option for studying the folding of small proteins (28).

**Constructing putative TS conformations.** Structures were constructed via constrained unfolding simulations (7) from the native PDB structure (1RIS). A common interpretation of $\phi^{exp}$ is the fraction of native contacts made by a particular residue in the TSE. We define a simulation $\phi$-value ($\phi^{sim}$), which may be calculated for any residue, as the fraction of native contacts made by residue $i$ in conformation $k$.

$$\phi_i^{sim} = \frac{N_i^k}{N_i^{native}} \tag{1}$$

Harmonic restraints were introduced using $\phi^{sim}$ to generate structures meeting experimental constraints (i.e having a $\phi^{sim} = \phi^{exp}$). This was accomplished by adding the $\phi$ energy, $E^\phi$, to the Go energy, $E^{Go}$.

$$E^{Tot} = E^{Go} + \Lambda \cdot E^\phi \tag{2}$$

$$E^\phi = \sum_{i=1}^{N}(\phi_i^{sim} - \phi_i^{exp})^2 \tag{3}$$

where $\Lambda$ is weighting factor, $i$ is the residue to be restrained, and $N$ is the total number of restraints. Experimental $\phi$-values used as restraints are summarized in Table 1. Unfolding simulations, using the energy function in Eqations 2-3, were initiated from the native state and propagated for $10^7$ MC steps. $\Lambda = 10^4$ ensured that restraints were quickly and fully met. The average $|\phi_i^{sim} - \phi_i^{restraint}|$ was less than $10^{-3}$ at the end of the minimization. Although Eq. 1 is not exact, it has been shown in several studies to be a reasonable proxy for $\phi^{exp}$ (2-6). Moreover, it has been shown that $\phi$-values predicted using our model and Eq. 1 correlate well with reliable experimental values (R = 0.70) (7).

**Verifying the TS.** A structure may not be a priori assumed a member of the TSE simply because it meet a set of $\phi$ restraints. To determine if a structure is a member of the TSE in a theoretically sound manner, one must measure its $p_{fold}$. A transmission coefficient, or $p_{fold}$ of 0.5 defines the TS. In order to determine this probability, 20 trial folding runs were performed for each structure. Because the calculation of $p_{fold}$ amounts to a Bernoulli trial, the relative error resulting from using $N$ runs scales as $N^{-1/2}$. Thus, the 20 trial folding runs estimate $p_{fold}$ to within 22% of the mean. As previously shown for protein folding (7) and many other complex systems (29), it is only necessary for each trial run to be long enough to confirm *commitment*. In order words, a particular structure is committed to folding if it is past the TS barrier and that the likelihood of re-crossing the barrier (i.e. unfolding) is negligible. As in earlier studies (14), we define $\tau_{commit}$ (the "commitment time" and duration of a $p_{fold}$ simulation) as $10^7$ MC steps. Commitment to folding was defined as a structure meeting (i) $E^{Go} < -1100$ (native $E^{Go} = -1569$), (ii) Backbone dRMS < 2.5Å, and (iii) fraction of native contacts ($Q$) > 0.75. $P_{fold}$ was then calculated as the fraction of the (20 total) runs that ended in a committed conformation. Commitment results from specific collapse, which is equivalent to complete nucleus formation in our model. After chain collapse under folding conditions (T = 1), the backbone never assumes an extended conformation. This is supported by extensive empirical observations in the simulated folding of crambin (15), protein G (14), and S6, which demonstrate that committed structures always proceed to the native state. Therefore, upon collapse, the protein is committed and one need only monitor (i) E to rule out side chain packing traps, (ii) compactness via structural parameters such as dRMS, and (iii) secondary structure formation ($Q$) to rule out folding traps.

**Results and Discussion**
**A refined picture of the S6 TSE.** The properties of our putative TSE are summarized in Table 2. The majority of conformations have intermediate $p_{fold}$ values, indicating the TSE was effectively sampled and identified. However, as previously observed (7), not all structures that meet the $\phi$ restraints are true TS conformations. This is not surprising, especially in consideration of the informative analogy to determination of a protein structure from NMR, which depends greatly on the number and quality of NOE restraints and the results of which often require additional refinement (30).

Figure 1 presents a side-by-side comparison of native S6 and the conformations in its TSE (i.e. with $p_{fold}$ = 0.5). In the TSE, the helix 1 is near native and the second and third β strands have well-formed secondary structure. Helix 2 makes a small number of native contacts, whereas strand 1 makes very few and strand 4 is nearly denatured. In the sense that the TSE exhibits an overall expanded native-like topology, one may interpret the predominance of intermediate $\phi$-values as signifying a diffuse TSE. However, as will be discussed next, a residue-level examination reveals a well-defined, fully established network of contacts comprising the specific nucleus.

**Nucleating contacts.** From a macroscopic perspective, experiment provides the picture of a diffuse, native-like TSE in S6, similar to that of CI2 (31). Although this view is consistent with nucleation-condensation behavior, deeper examination is required to understand the formation and role of the specific nucleus of non-local contacts that define the TSE topology. Here, we define the fully formed (in all TSE conformations) subset of long range contacts as the specific nucleus. Identifying the specific nucleus is important since after passing the energetic barrier to nucleation a structure is committed to rapid backbone folding followed by a slower side chain relaxation process (32). To determine the role of individual residues in the nucleation process, contact maps were created for the native- and the TSE conformations (Figure 2 a and b). From these contact maps, it is clear that a small number of residues make most of the long-range contacts. These non-local contacts form between strands 1 and 3, strands 2 and 3, helix 2 and strand 1, and to a lesser degree between the helix 1-strand2 loop and second helix.

Figure 3 shows the average total non-local contacts made by each residue in the TSE (unfolding restraints determine the number of contacts made by restrained residues and indirectly influence neighboring contacts), clarifying the role of individual residues in nucleation. Several residues with high $\phi^{exp}$, such as I26, L30, and M67, do not make a large number of long-range contacts. However, I26 and L30 make the only long-distance contacts in helix 1, which appears to form early in folding and which contacts I8, V65, and L75 in the TSE.

All other residues with high $\phi^{exp}$ made a large number of non-local contacts (Y4, V6, I8, F60, V65, and L75). Additionally, every residue with $\phi^{exp} < 0.20$ made few (< 5) non-local contacts. Three residues, which stand out in Figure 3b for their large number of non-local contacts, cannot be reliably characterized by protein engineering experiments due to insufficient stability changes. Y33 ($\phi^{sim} = 0.40$), which is located at the end of helix 1 and interacts with the high $\phi$ residues L30 and L75 to bring together helix 1 and 2 and form part of the hydrophobic core, could not yield a $\phi^{exp}$ because the Y33A mutation only destabilized the protein by 0.38 kcal/mol. L48 ($\phi^{sim} = 0.74$) anchors the long loop region between sheets 2 and 3 by forming a buried, hydrophobic cluster with I52 and F60 ($\phi^{exp} = 0.36$). Interestingly, truncation of residues L48 and I52 to A, which show strong conservation in the split β-α-β family (11), do not result in significant destabilization of the protein.

The important role of I26 and L30 in defining the topological orientation of helix 1 in the folding nucleus is also indicated experimentally through kinetic $m$-values, (Figure 4). Truncation of the I26 and L30 side chains produce an increase of the unfolding $m$-value, $m_u$ (11), indicative of a pathway shift along secondary order parameters (33). If helix 1 is not anchored by long-range contacts early in folding, S6 folds through an alternative "mutant" trajectory. Consistently, the effect is enhanced in the double mutants I26A/L30A ($m_u = 0.90$), I8A/I26A ($m_u = 0.75$) and L30A/L75A ($m_u = 0.85$) that exhibit unfolding $m$-values nearly twice as high as for the wild-type protein ($m_u = 0.50$) (Figure 4). The ability of S6 to fold through such an alternative nucleus has been demonstrated independently by circular permutation experiments (12). Interestingly, I26A/L79A also has a high $m_u$ (0.75), which is higher than I26A alone ($m_u = 0.65$), i.e. the change is additive for L79. However, double mutants at other positions in S6 show wild-type level or lower $m_u$ values (i.e. show various degree of Hammond shift). Moreover, we see that double mutants 26/79, 8/26, 30/75 and 26/30, and possibly the single mutants L30, W62, Y63 and V65 show higher values of the refolding $m_f$, indicating that the network of contacts defined by these positions is to some extent present in the denatured ensemble.

**Evolutionary signals of nucleation.** Evolutionary signals of universally conserved residue positions have previously been linked to information about the stability, kinetics, and function of proteins. To this end, we analyzed proteins which share the S6 fold, but not its sequence, through "Conservation of Conservation" (CoC) (34). This method has also been shown effective at identifying protein folding nuclei (35). For S6, 42 FSSP families with Dali Z-score greater than four and sequence ID less than 25% were considered (36). The results of this analysis are presented in Figure 3c, where low values indicate a position's high conservation within the studied families.

The first cluster of low CoC residues are located in sheet 1 and include V6, I8, L10, and P12. V6 and I8 have high measured $\phi^{exp}$ and make a large number of non-local contacts in the TSE, as does P12. In contrast, L10, which is part of this hydrophobic cluster, makes very few non-local contacts, which may be an artifact of neighboring V9's low ($\phi^{exp} = 0.07$) restraint. The next two low CoC residues, I26 and L30 in helix 1, exhibit relatively high $\phi^{exp}$. Although these residues make comparatively few long-distance contacts, they play an unambiguous, important role as the only non-local contacts made by the early-forming first helix. V37, which exhibits low CoC, is involved in tertiary interactions in the second strand. Although it exhibits a moderate $\phi^{exp} = 0.24$, V37 has been suggested to play a role late in the TSE (11). Mutation of the low CoC residue I52, which is located in a hydrophobic cluster in the long loop region connecting strands 2 and 3 and which makes a large number of non-local contacts in the TSE, did not result in significant destabilization so a $\phi$-value could not be measured. Residues L48, I52, and F60 are in close contact with R87, which may play a functional role in RNA binding (9). In fact, the majority of residues in this loop region exhibit a high degree of order in the TSE (Figure 3a), despite making few long-range contacts. Loop folding, which appears to occur semi-autonomously of nucleation, may be a result of the protein-binding functional role required for ribosome formation (11). The low CoC residues Y63 and V65 in strand 3 are located at the center of the high-$\phi$ cluster that includes F60 and M67. Although Y63 has a lower $\phi^{exp}$, it makes far more long-range contacts in the TSE than the higher $\phi^{exp}$ V65 or M67. In this sense, experiment may belie the importance of Y63's role in nucleation. Another group of low CoC residues is V72, L75, and L79. L75 has a high $\phi^{exp}$ and V72 exhibits the highest $\phi^{sim}$ value in helix 2. L79, clearly important in non-local nucleation contacts, was restrained by a $\phi^{exp} = 0.16$, which either indicates a minor role in the TSE or, more likely, a role which was mitigated by an artificially low restraint.

Fig. 3c indicates that CoC is strongly related to nucleation: e.g. out of 14 residues implicated in nucleation by φ-value, $p_{fold}$ and m-value analyses, 12 have low CoC < 0.5. Probability of that occurring by chance is approximately $10^{-7}$. The value of CoC clearly lies in the way in which it complements $\phi^{exp}$ and simulation. To identify and understand the folding nucleus, simple examination of $\phi^{exp}$ is insufficient. One such example, L79, has a low $\phi^{exp}$ but plays a role in nucleation as suggested by simulations and m-value analysis. However the role of L79 may be underestimated in φ-value analysis because of denatured state contacts. From the above experimental, simulation, and evolutionary data, we gain a clear and mutually consistent picture of the TSE, which allows us to examine the ordered sequence of events before and after nucleation and thereby fully describe the process of S6 folding.

**Capturing the uphill and downhill events in folding.** $P_{fold}$ analysis allows each conformation to be classified as a pre-, post-, or TS structure, from which we may infer the order of events in folding. To this end, contact statistics were separately collected and normalized for $p_{fold} < 0.4$ and $> 0.6$ (pre- and post-TSE) structures. To determine the contacts prevalent in each state, and to remove the influence of contacts that are common to all states, a difference matrix of contacts (post – pre-TSE) was calculated and plotted (Figure 2c). The results of this analysis may be compared with the results of continuous $\phi$-value experiments, which probe the order of folding events by shifting the TSE by perturbing protein stability (37). These $\phi$-value changes are rigorously understood by the Hammond postulate (38), which is readily applied to proteins (39, 40). Such an analysis was first carried out on U1A (37), and later conducted on CI2 (35) and S6 (11), by measuring shifts in kinetic $m$ and $\phi$-values in response to changes in denaturant (guanidinium chloride, GdmCl) concentration.

It has been suggested that mutations that decrease $m_u$, and consequently exhibit higher $\phi$-values at 7M GdmCl, make their major energetic contribution after the folding barrier (11). In S6, such residues (Y4, V37, V88, V90) indicate increasing structure in helix 2 and the C-terminal strand after nucleation. The same trend is reflected in Figure 2c. Experiment also suggests that formation of the interface between the N and C termini (high $\phi$-values at 7M GdmCl for Y4, V88, and V90) occurs late in folding. Our results show the N and C termini and the contacts between strands 1 and 3 consolidate mainly after nucleation. At the transition midpoint, experiment suggests that strands 2 and 3 come together, which is also observed in simulation. In fact, different contacts in strands 2 and 3 are observed in the pre- and post-TSE structures, indicating that this process spans the transition region. This is indicated by Figure 2, which shows the structured strand 2-3 region. Figure 3b shows the critical role of residues making non-local contacts, including those made by L48, F60 Q57, and W62, which were not identified by experiment. The docking of strand 3 with helix 2 appears important in nucleation both from high $\phi^{exp}$ (V65, M67, and L75) and the large number of non-local contacts in the TS (especially L75). As previously suggested (11), residues with increased $m_f$ values, appear involved in structure before the transition barrier in our simulations. These positions (V6, L19, L30, L61, and L65) are in helix 1 and strand 3.

Because pre- and post-TS conformations are subject to the same $\phi^{exp}$ restraints as the TSE, they do not differ markedly in their number of contacts, $Rg$, or dRMS (Table 2), as might be expected for structures earlier or later in folding. Although $\phi^{exp}$ suggests an overall more expanded TS at 0M than at 7M GdmCl, for the above reasons our simulation does not. When $\phi^{0M}$ and $\phi^{7M}$ were used in simulation, they resulted in ensembles with $p_{fold} \approx 0$ and $\approx 1$, respectively. Although $\phi$ values provide a measure of TS structure, due to the differences in experimental folding conditions (0 or 7M), one should not expect the same TSE. Indeed, due to differences in $\phi$ values, we know the TSEs probed under different conditions are dissimilar. Thus it is logical that these TSEs exhibit different $p_{fold}$ when refolded under different conditions. Nevertheless, because $p_{fold}$ is a good reaction coordinate for protein folding (29), we are able to learn the ordered details of folding from structures derived from $\phi^{midpoint}$. This demonstrates the dominant role and robustness of nucleating contacts and the overall plasticity of other contacts. Comparison of the interpretations from $m_f$, $m_u$, $\phi^{0M}$, and $\phi^{7M}$ with our pre- and post-TS conformations, support the interpretation of solvent manipulation as an effective way to probe shifts in the TSE and hence the order of events in folding.

**The synergy of simulation, experiment, and evolutionary data yield a deeper understanding of S6 folding.** Through simulation we have built an atomic-level picture of the nucleating contacts and the order of folding events in S6, which is consistent with, and extends the interpretation of an extensive body of evolutionary and experimental data. S6 folds through nucleation-condensation mechanism with strand 1 and helix 1 coming together via long-range contacts within the V6-I8-I26-L30 cluster. This specific nucleus is stabilized by V65 and L75 in strand 3 and helix 2. Most other residues that appear important in only one or two of the above methods are located in the hydrophobic core contacting (L10, P12, Y33, V37, Y63, M67, and V72). There is also an interesting group of three residues in the large strand 2-3 loop. The high-$\phi$, high number of non-local contacts residue F60 is flanked by L48 and I52, which appear to anchor the loop region through late forming contacts with V9. This loop plays an important role in ribosome formation, where it binds protein S18 (10) and may thus benefit functionally by being independent of the cooperative folding unit. Protein S6 represents an example of classic nucleation-condensation behavior (18). Upon the formation of a specific nucleus (the entropic barrier), folding proceeds downhill energetically through the condensation of a plastic, hydrophobic cloud of supporting contacts.

The order of folding events suggested by experiment is made clear through simulation. Contacts form first between strand 1 and helix 1, with nucleation centering on the formation of the V6-I8-I26-L30 cluster. V65 and L75, which are separated from this cluster in sequence but close in structure, also appear involved in the nucleus. While experiment has not yet determined the degree of secondary structure in helix 1, simulation shows it is well formed prior to nucleation. The large loop separating strands 2 and 3 closes in around F60, L48, and I52, docking with V9 after nucleation. Helix 2 forms secondary structure and docks to strand 1 largely after nucleation. Strands 1 and 4 also come together after nucleation. In addition to being entropically disfavored, these late-forming contacts interact with both RNA and protein S18 in the ribosome (10) and may gain a functional advantage from such conformational flexibility.

As we observe in $p_{fold}$ simulations, upon nucleation, if a structure moves beyond the transition barrier, it is committed to complete and rapid folding. Although most $\phi^{exp}$ have intermediate values in S6, suggesting a TS that is uniformly collapsed, TS conformations exhibit distinct regions of order and disorder in their secondary and tertiary structure. Decreased experimental $m_u$ values around the equilibrium transition region suggest the possibility of partial unfolding of the native state (11). Consistently, our simulations suggest that such fluctuations occur outside the cooperative unit, in strands 1 and 4 and in the strand 2-3 loop. In this sense, one could describe the TSE as "diffuse" in that it exhibits a largely collapsed backbone. However, it is not simply an expanded version of the native topology. Specific nucleation defines the topology of the TSE and establishes the framework for critical contact formation. After formation of this cooperative unit, a structure is committed to folding.

The specific nucleus is a fixed set of contacts common to all TSE conformations. Other supporting contacts within the critical nucleus are, to a certain extent, heterogeneous. Whereas some of these supporting contacts may be formed in one subset of the TSE ($p_{fold}$ = 0.5) conformations, other subsets of TSE conformations may feature different supporting contacts. In this sense the heterogeneous cloud of hydrophobic contacts that support the nucleus is variable or plastic. There is also a plasticity of the contacts outside the cooperative unit (strand 2-3 loop region, for example) that, in principle, may be reached by ground (native) state fluctuations or local unfolding. These regions represent deviations from global cooperativity, but not from thermodynamic two-state behavior. This observed plasticity outside the cooperative unit may be tentatively ascribed to the function of RNA and protein binding in S6. However, similar regions in other proteins may be implicated in erroneous side reactions such as aggregation.

**Conclusions.** We have built an atom-level model of the TSE and evolutionary CoC analysis, to obtain a deeper understanding of folding by probing the role of residues which are inaccessible by experiment. The order of folding events we infer from simulations correlates with and provides support for the interpretation of experiments that change $\phi$-values by modulating [GdmCl]. These data also present strong evidence that the mechanistic details in our MC-Go model are realistic and thus allow for a detailed structural interpretation of $\phi^{exp}$. Simple examination of $\phi^{exp}$ is insufficient for fully understanding the nucleation and folding of S6 or other proteins. It is only thought the synergy of experiment and simulation that a detailed and complete model may be built. Most importantly, we observe that as long as the cooperative unit, the folding nucleus, is formed in the TSE, other parts of structure may fluctuate. In this regard, conformations along folding pathway display structural plasticity. For S6, which binds to RNA and protein subsequent to folding, this plasticity may be a direct consequence of the conformational flexibility required to achieve biological function.

**Acknowledgements**
We thank Eric J. Deeds and Jason E. Donald for their comments on the manuscript. This work was supported by an HHMI pre-doctoral fellowship (IAH) and by NIH ROI 52176.

**Figure Legends**

**Figure 1.** Two views of S6 native (a) and representative transition state (b) structures. Nucleus residues (V6, I8, I26, L30, P60, V65, L75) are shown explicitly. Progression along the chain is indicated by color, from blue (N-terminus) to red (C-terminus). Images were created using iMol.

**Figure 2.** 1RIS contact matrices for the (a) native, (b) TSE, and (c) the difference matrix (Post – Pre TS) with the upper left corresponding to positive values and the bottom left corresponding to negative values.

**Figure 3.** S6 TSE contact statistics. (a) The average fraction of native contacts made by each reside. Circles indicate $\phi$-restraints. (b) The average number of non-local (excluding n to n±5) contacts made by each reside. (c) Conservation of Conservation in S6. Residues participating in nucleation (as determined by $p_{fold}$ and m-value analysis) are indicated by an x and residues with $\phi^{exp} \geq 0.3$ are circled, highlighting agreement between CoC and $\phi^{exp}$. Additionally, CoC (along with $m$-values and simulation) point out to some nucleus residues with apparently low $\phi^{exp}$, such as L79. CoC may also identify residues with functional importance, such as I52.

**Figure 4.** Experimental $m_u$ and $m_f$ values for wild type and mutant S6. The $m$-values were determined by standard procedures from the linear regions of the chevron plots, excluding regions of downward curvature that are sometimes seen at high denaturant concentrations (11). Residues in the specific nucleus display the highest $m$-values, indicative of alterations of the folding trajectory and loss of residual structure in the denatured ensemble for $m_u$ and $m_f$, respectively.

**Table 1.** S6 $\phi$-values from experiment (11) and simulation, along with CoC value and average number of non-local contacts ($N_{nonlocal}$) in the TSE. Mutations and residue numbers are listed in the left column and the secondary structure element to which each residue belongs is indicated in the right column.

**Table 2.** Ensemble properties. Averages and standard deviations were calculated for the entire putative TSE and for subsets. For the native conformation, N = 1569 and $Rg$ = 12.95Å.

**Figure 1**

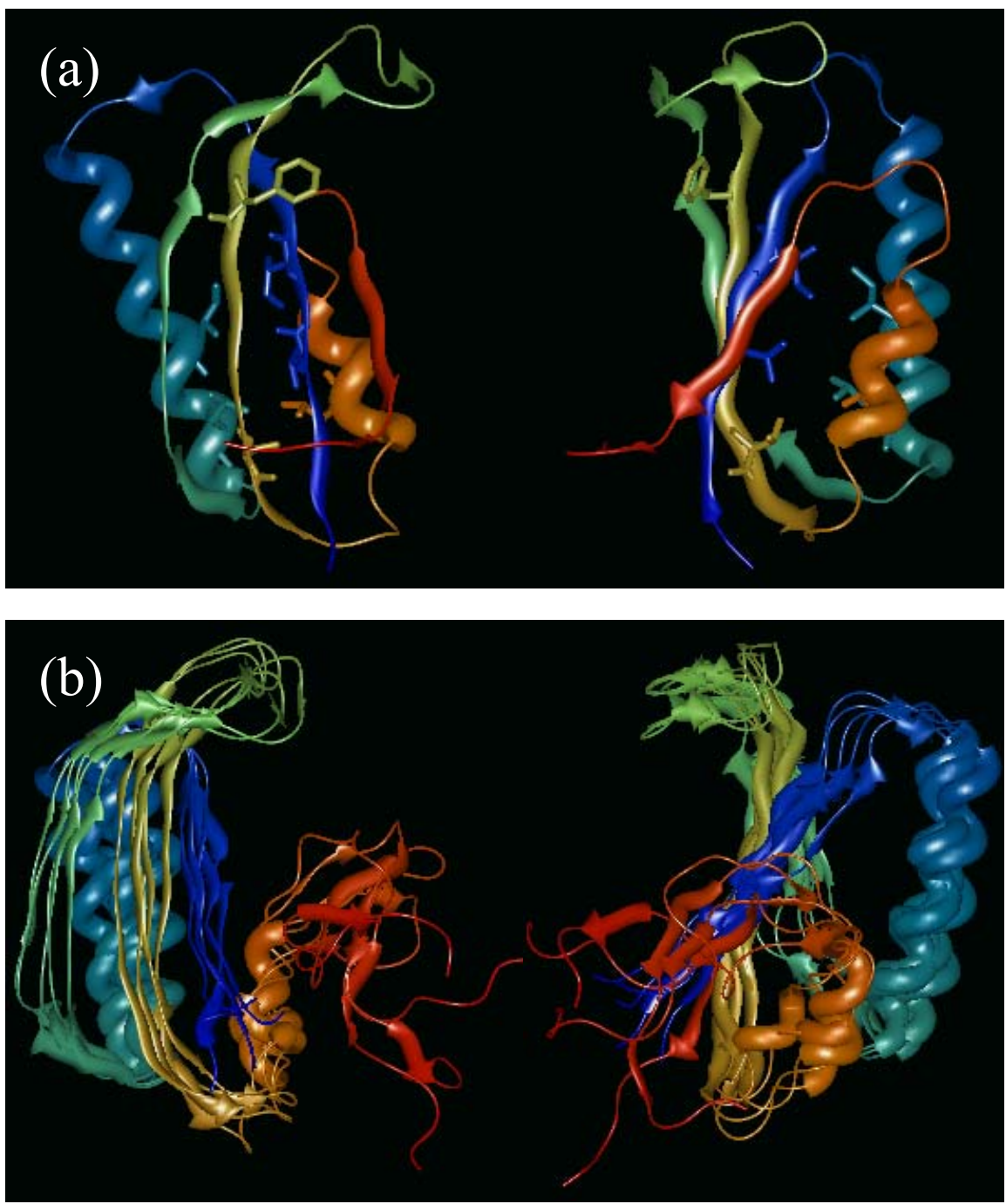

**Figure 2**

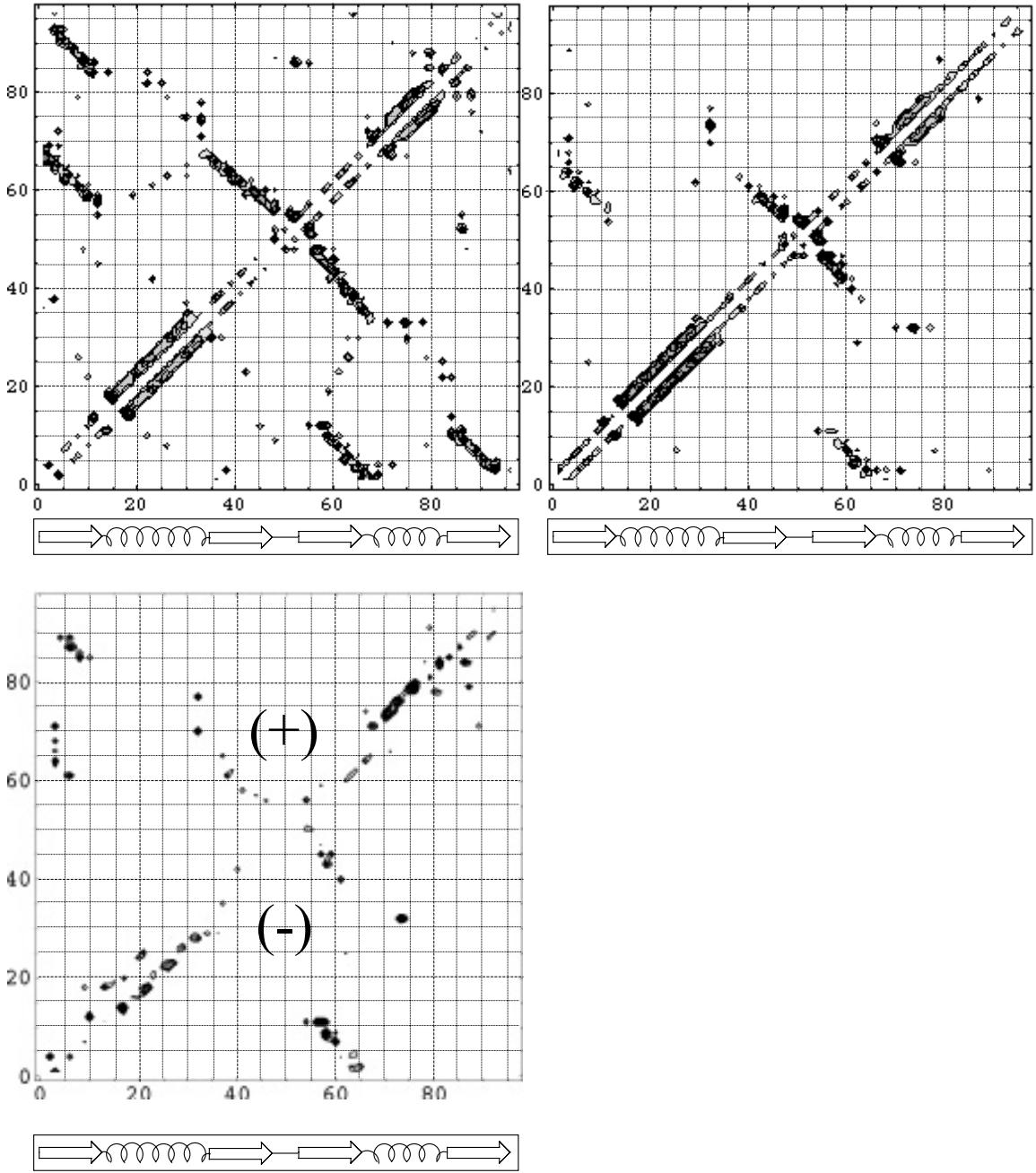

**Figure 3**

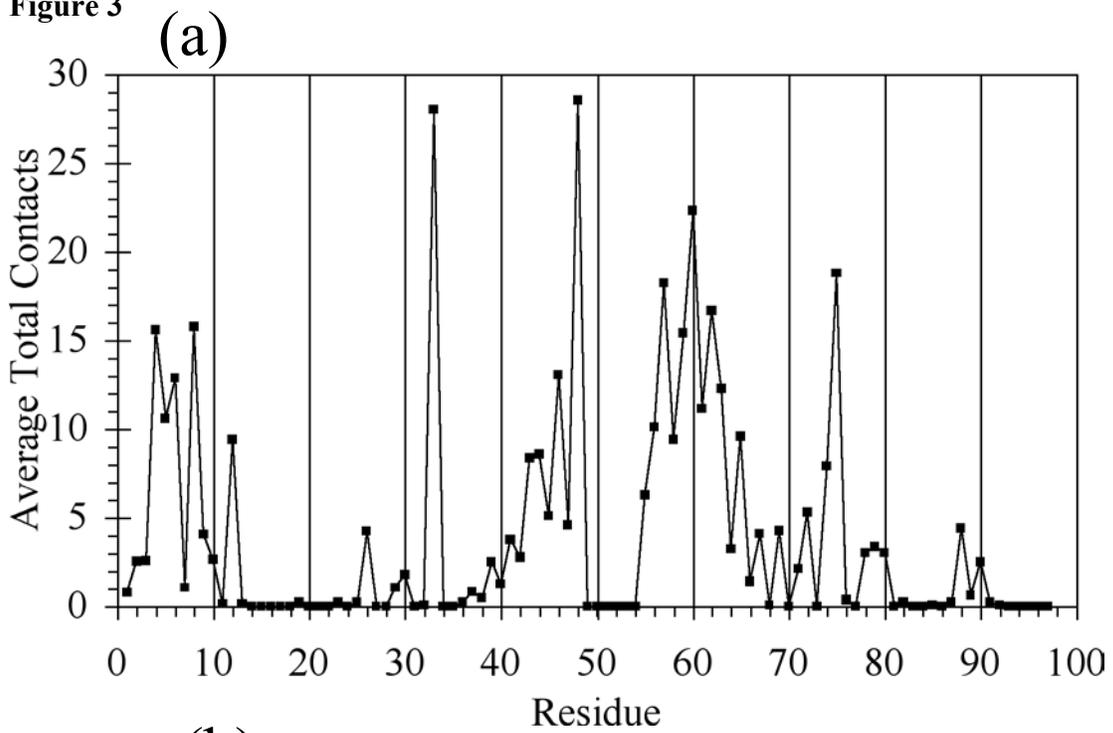

(a)

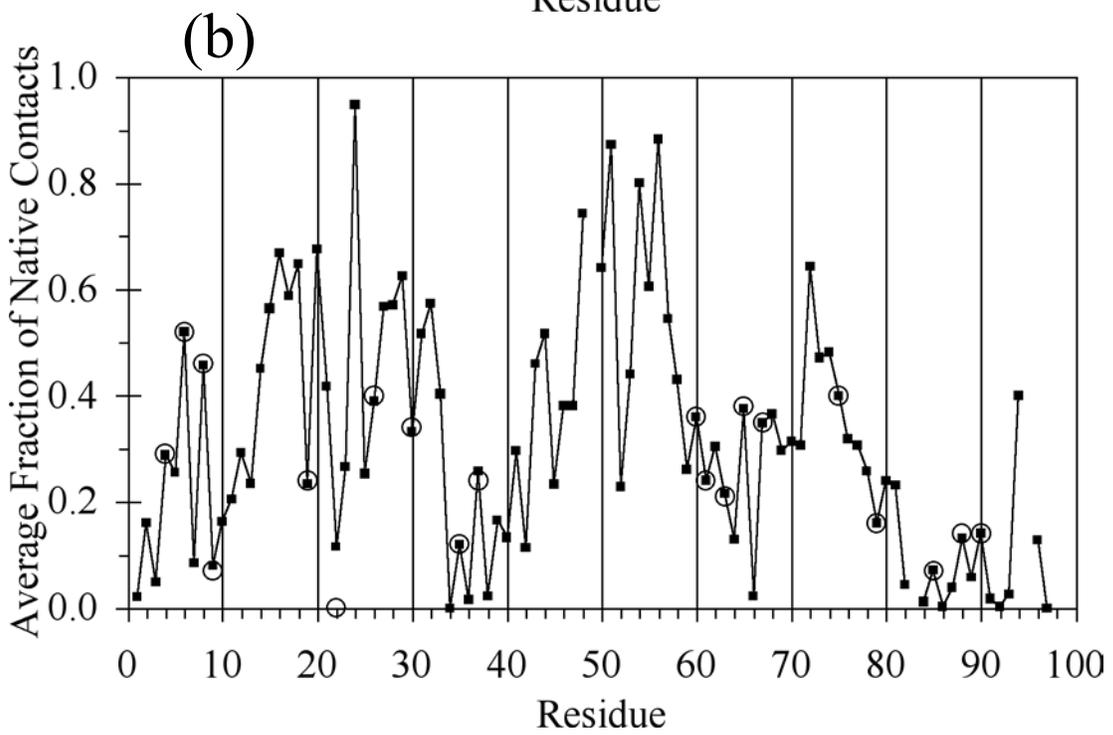

(b)

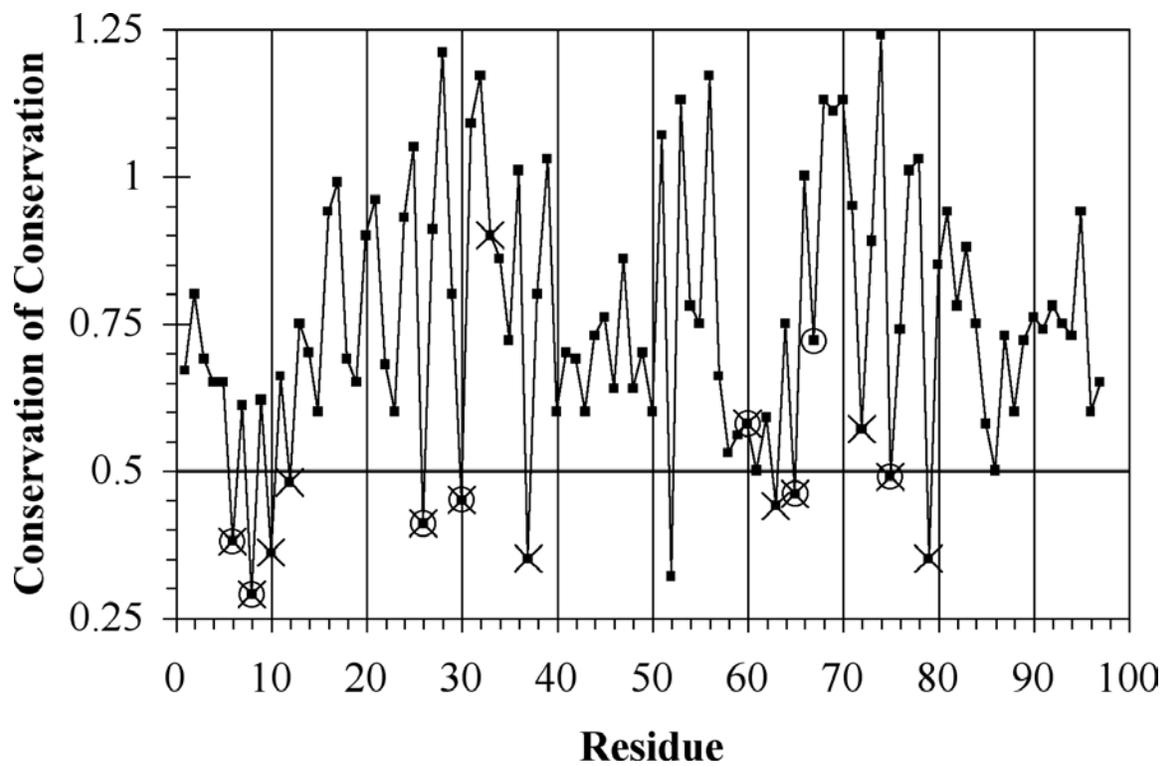

**Figure 4**

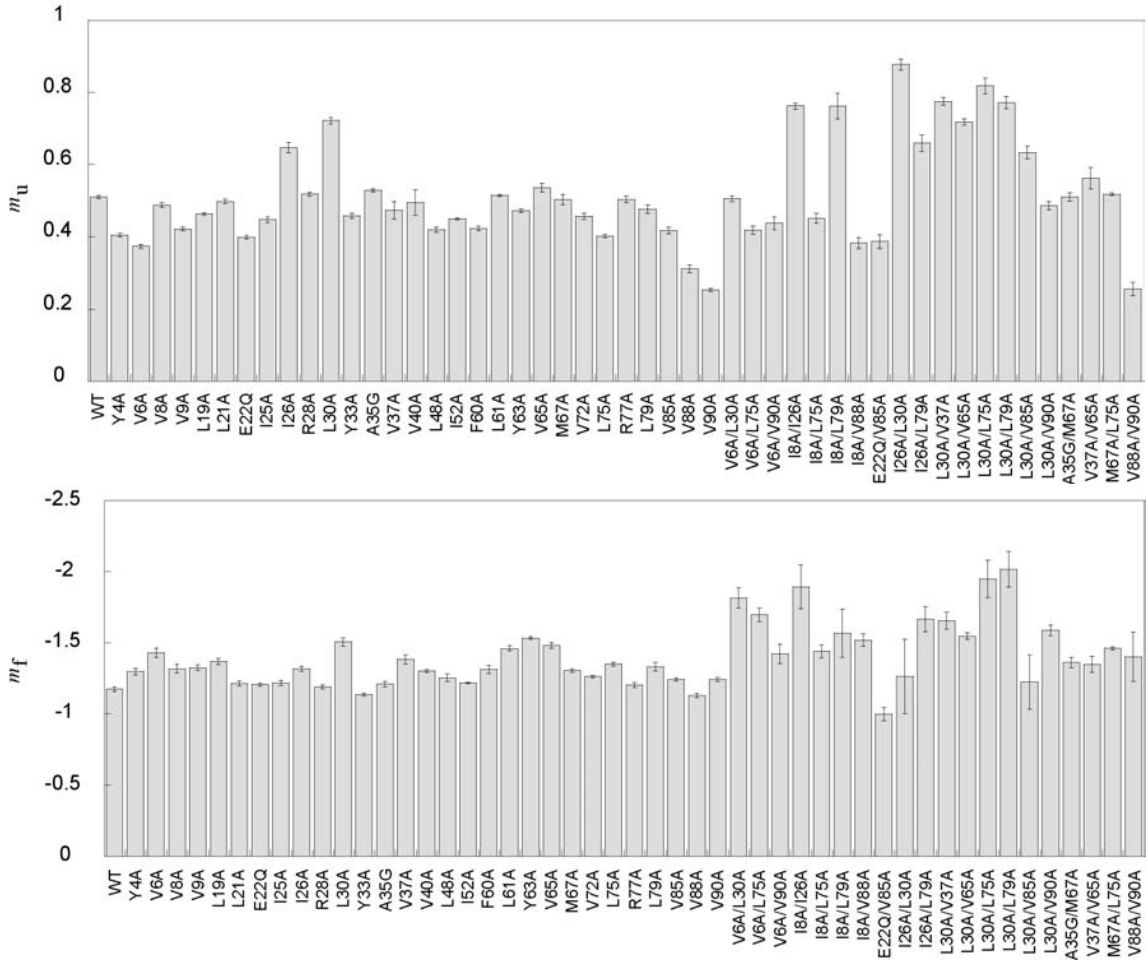

**Table 1**

| Mutant | $\phi^{midpoint}$ | $\phi^{sim}$ | CoC | $N_{nonlocal}$ | |
|---|---|---|---|---|---|
| Y4A | 0.29 | 0.29 | 0.65 | 15.57 | β1 |
| V6A | 0.52 | 0.52 | 0.38 | 12.87 | |
| I8A | 0.46 | 0.46 | 0.29 | 15.74 | |
| V9A | 0.07 | 0.08 | 0.62 | 4.04 | |
| L10 | - | 0.16 | 0.36 | 2.61 | |
| P12 | - | 0.29 | 0.48 | 9.39 | loop |
| L19A | 0.24 | 0.23 | 0.65 | 0.22 | α1 |
| E22N | 0.00 | 0.12 | 0.68 | 0.00 | |
| I26A | 0.40 | 0.39 | 0.41 | 4.22 | |
| L30A | 0.34 | 0.33 | 0.45 | 1.74 | |
| Y33 | - | 0.40 | 0.90 | 28.00 | loop |
| A35G | 0.12 | 0.12 | 0.72 | 0.00 | |
| V37A | 0.24 | 0.26 | 0.35 | 0.83 | β2 |
| L48 | - | 0.74 | 0.64 | 28.52 | |
| I52 | - | 0.23 | 0.32 | 0.00 | loop |
| Q57 | - | 0.54 | 0.66 | 18.22 | |
| F60A | 0.36 | 0.36 | 0.58 | 22.30 | β3 |
| L61A | 0.24 | 0.24 | 0.50 | 11.13 | |
| W62 | - | 0.30 | 0.59 | 16.65 | |
| Y63A | 0.21 | 0.22 | 0.44 | 12.26 | |
| V65A | 0.38 | 0.37 | 0.46 | 9.57 | |
| M67A | 0.35 | 0.35 | 0.72 | 4.09 | |
| V72 | - | 0.64 | 0.57 | 5.30 | α2 |
| L75A | 0.40 | 0.40 | 0.49 | 18.78 | |
| L79A | 0.16 | 0.16 | 0.35 | 3.35 | |
| V85A | 0.07 | 0.07 | 0.58 | 0.04 | loop |
| V88A | 0.14 | 0.13 | 0.60 | 4.39 | β4 |
| V90A | 0.14 | 0.14 | 0.76 | 2.48 | |

**Table 2**

| Ensemble | N | $p_{fold}$ | $N^{nat}$ | $N^{nnat}$ | Rg | dRMS |
|---|---|---|---|---|---|---|
| All | 90 | 0.35±0.23 | 439±29 | 229±29 | 15.0±0.5 | 5.0±0.9 |
| Pre-TS | 56 | 0.20±0.11 | 433±25 | 226±24 | 15.2±0.6 | 5.2±1.0 |
| True TS | 23 | 0.52±0.07 | 452±32 | 229±32 | 14.7±0.4 | 4.6±0.07 |
| Post-TS | 11 | 0.77±0.06 | 438±37 | 245±40 | 14.9±0.03 | 4.6±0.06 |